\documentclass[aps,prm,reprint,superscriptaddress,floatfix]{revtex4-2}

\usepackage{graphicx}
\usepackage{amssymb,amsmath}
\usepackage{dcolumn}
\usepackage{bm}
\usepackage[mathlines]{lineno}
\usepackage{multirow}
\usepackage{color}
\usepackage{threeparttable}
\usepackage{booktabs}
\usepackage[colorlinks,linkcolor=blue,anchorcolor=blue,citecolor=blue,urlcolor=blue]{hyperref}
 \newcommand{\modR}[1] {{#1}}

\begin{document}

\title{Low-symmetry two-dimensional BNP$_2$ and C$_2$SiS structures\\
       with high and anisotropic carrier mobilities}

\author{Shixin Song}
\affiliation{School of Physics,
             Southeast University,
             Nanjing, 211189, PRC}

\author{Jie Guan}
\email
            {guanjie@seu.edu.cn}%
\affiliation{School of Physics,
             Southeast University,
             Nanjing, 211189, PRC}

\author{David Tom\'{a}nek}
\email
            {tomanek@msu.edu}%
\affiliation{Physics and Astronomy Department,
             Michigan State University,
             East Lansing, Michigan 48824, USA}

\begin{abstract}
We study the stability and electronic structure of previously
unexplored two-dimensional (2D) ternary compounds BNP$_2$ and
C$_2$SiS. Using {\em ab initio} density functional theory, we have
identified four stable allotropes of each ternary compound and
confirmed their stability by calculated phonon spectra and
molecular dynamics simulations. Whereas all BNP$_2$ allotropes are
semiconducting, we find C$_2$SiS, depending on the allotrope, to
be semiconducting or semimetallic. The fundamental band gaps of
the semiconducting allotropes we study range from $1.4$~eV to
$2.2$~eV at the HSE06 level
and display carrier mobilities as high as
$1.5{\times}10^5$~cm$^2$V$^{-1}$s$^{-1}$. Such high mobilities are
quite uncommon in semiconductors with so wide band gaps.
Structural ridges in the geometry of all allotropes cause a high
anisotropy in their mechanical and transport properties, promising
a wide range of applications in electronics and optoelectronics.
\end{abstract}


\maketitle

\section*{Introduction}

Two-dimensional (2D) materials have intrigued researchers around
the world since the successful mechanical exfoliation of
graphene~\cite{Novoselov04S}. %
\modR{ %
Even though graphene remains unsurpassed in terms of high charge
carrier mobility, %
}%
the vanishing band gap in the pristine material precludes its use
from semiconducting circuitry~\cite{{Novoselov2005},{Zhang2005}}.
Other 2D materials including transition metal dichalchogenides
(TMDs) such as MoS$_2$ have sizable band gaps, but are not as
useful due to their low carrier
mobility~\cite{{Kis2011},{Fuhrer2013}}. Phosphorene, a monolayer
of black phosphorus, combines high and anisotropic carrier
mobility with a sizeable and tunable band
gap%
\modR{%
~\cite{{DT229},{Li2014},{WangL2015}},
} %
but is unstable under ambient conditions~\cite{Hersam14}. In spite
of significant efforts to improve the performance of 2D materials
in semiconducting devices, the progress has been moderate. There
is a need to find new 2D semiconductors with substantial band gaps
and high carrier
\modR{ %
mobilities.%
}%

Phosphorus carbide (PC), a recently proposed 2D material, has been
predicted to be stable~\cite{2DPC16} and to display promising
electronic behavior including high carrier
mobility~\cite{Wang16PC}. Of the stable allotropes, the
semiconducting $\alpha_1$-PC phase, also called black phosphorus
carbide ($b$-PC), has been successfully synthesized. It shows
\modR{ %
a %
}%
high field-effect mobility $\mu=1995$~cm$^2$V$^{-1}$s$^{-1}$ of
holes at room temperature~\cite{Tan17AM}, good infrared
response~\cite{Tan18AM} and tunable anisotropic plasmonic
performance~\cite{PCplasmon}. The narrow band gap, however, limits
the performance of PC-based field-effect transistors (FETs) due to
a relatively low ON/OFF ratio~\cite{Tan17AM}. Experimental data
for 2D GeP~\cite{2dGeP}, another low-symmetry IV-V compound with
strongly anisotropic conductance, indicate a high field-effect
ON/OFF ratio ${\approx}1\times10^4$, but a low carrier mobility
${\mu}=0.35$~cm$^2$V$^{-1}$s$^{-1}$. Theoretical explorations have
been extended to other 2D IV-V compounds, which are not
isoelectronic to PC. These include %
germanium triphosphide~\cite{Jing17GeP3} GeP$_3$, %
tin triphosphide~\cite{SnP3} SnP$_3$, and %
phosphorus hexacarbide~\cite{PC6} PC$_6$. However, all these
systems have narrow band gaps similar to $\alpha_1$-PC and share a
hexagonal honeycomb lattice structure and thus a weak transport
anisotropy. 2D structures with a substantial band gap, high
carrier mobility and strong in-layer anisotropy are still missing.

In search of such 2D materials inspired by anisotropic 2D PC
structures, we have applied an effective design strategy known as
``isoelectronic substitution''. This process involves substituting
certain elements in the structure by their neighbors in the
periodic table, yet keeping the total valence electron count
unchanged. This approach, which has been successfully applied in
both 3D and 2D systems, allows to change physical and chemical
properties of the system without drastically changing the
structure. In this way, the diamond structure of bulk silicon with
a diatomic unit cell can be changed to the isoelectronic
Si$_3$AlP, when one Si atom is substituted by Al and the other
\modR{ %
by
}%
P in every other unit cell, thus significantly increasing light
absorption in the visible region~\cite{Si3AlP}. In semimetallic 2D
graphene with a diatomic basis, substituting one C atom by B and
the other by N forms the $h$-BN structure, a wide-gap insulator.
In a similar way, substituting every other atom in phosphorene by
Si and S atoms results in the 2D SiS
structure~\cite{{DT247},{yang16NL}}. The same isoelectronic
substitution in 2D group V systems leads to 2D group IV-VI
compounds including GeS, GeSe, and SnS with a lower symmetry and a
wide range of physical
properties~\cite{{Gomes15prb},{GeS15prb},{Fei16prl},{Haleoot17prl}}.

\begin{figure*}[t]
\includegraphics[width=1.8\columnwidth]{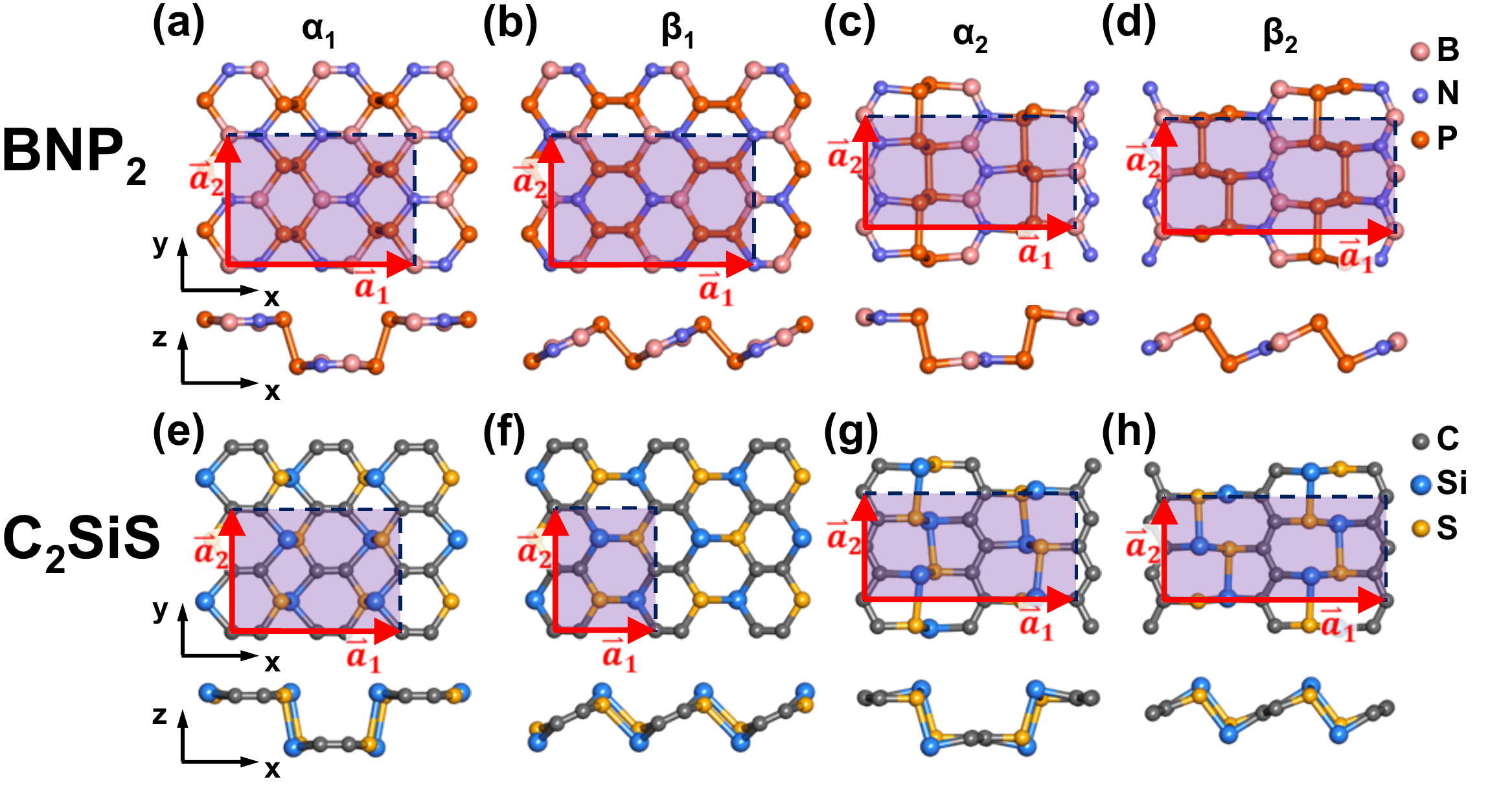}
\caption{%
Ball-and-stick models of relaxed monolayer structures of %
(a) $\alpha_1$, %
(b) $\beta_1$, %
(c) $\alpha_2$ and %
(d) $\beta_2$ %
    allotropes of BNP$_2$, and %
(e) $\alpha_1$, %
(f) $\beta_1$, %
(g) $\alpha_2$ and %
(h) $\beta_2$ %
    allotropes of C$_2$SiS %
in top and side view. The lattice vectors are indicated by red
arrows. %
\label{fig1}}
\end{figure*}


In this study, we %
\modR{%
propose isoelectronic substitution in 2D structures of phosphorus
carbide that leads to %
}%
previously unknown ternary compounds BNP$_2$ and
C$_2$SiS. Our {\em ab initio} density functional calculations
identify four stable allotropes of each compound that share the 2D
geometry with PC. Calculated phonon spectra and {\em ab initio}
molecular dynamics (MD) simulations confirm the stability of each
of these allotropes. Due to structural ridges in the geometry, all
allotropes considered exhibit a significant anisotropy in the
elastic response. Electronic structure calculations indicate that
$\alpha_1$ and $\beta_1$ phases of C$_2$SiS, as well as all four
BNP$_2$ allotropes, are semiconductors with a wide range of band
gap values. Carrier mobility calculations show that most of the
semiconducting allotropes exhibit high and strongly anisotropic
carrier mobilities. The remaining $\alpha_2$-C$_2$SiS and
$\beta_2$-C$_2$SiS allotropes were found to be semimetallic,
displaying an elliptically distorted Dirac cone in their band
structure caused by their structural anisotropy.

\begin{table*}[t]
\caption{Calculated equilibrium properties of BNP$_2$ and C$_2$SiS
         2D allotropes.} %
\setlength{\tabcolsep}{2.5mm}
\begin{tabular}{lcccccccc}
\hline\hline %
                               &\multicolumn{4}{c}{BNP$_2$} %
                               &\multicolumn{4}{c}{C$_2$SiS} \\%
                               & $\alpha_1$          & $\beta_1$ %
                               & $\alpha_2$          & $\beta_2$ %
                               & $\alpha_1$          & $\beta_1$ %
                               & $\alpha_2$          & $\beta_2$ \\%
\hline %
$|\vec{a}_1|$~({\AA})$^a$      & 8.50                & 9.39 %
                               & 9.44                & 10.44 %
                               & 8.07                & 4.75 %
                               & 9.83                & 10.61 \\%
$|\vec{a}_2|$~({\AA})$^a$      & 5.97                & 5.97 %
                               & 5.14                & 5.14 %
                               & 5.87                & 5.86 %
                               & 5.03                & 5.03 \\%
\hline %
$d_{P-P}$~(\AA)$^b$            & 2.25                & 2.25 %
                               & 2.27$^h$            & 2.27$^h$ %
                               &                     &     %
                               &                     &   \\%
                               &                     & %
                               & 2.24$^i$            & 2.23$^i$ %
                               &                     & %
                               &                     &   \\%
$d_{P-N}$~(\AA)$^b$            & 1.78                & 1.78 %
                               & 1.79                & 1.79 %
                               &                     &      %
                               &                     &    \\%
$d_{P-B}$~(\AA)$^b$            & 1.94                & 1.94 %
                               & 1.95                & 1.95 %
                               &                     &      %
                               &                     &    \\%
$d_{B-N}$~(\AA)$^b$            & 1.40                & 1.40 %
                               & 1.46                & 1.46 %
                               &                     &      %
                               &                     &    \\%
\hline %
$d_{Si-S}$~(\AA)$^b$           &                     &  %
                               &                     &  %
                               & 2.49                & 2.46 %
                               & 2.41$^h$            & 2.40$^h$ \\%
                               &                     & %
                               &                     & %
                               &                     & %
                               & 2.35$^i$            & 2.35$^i$ \\%
$d_{C-S}$~(\AA)$^b$            &                     &      %
                               &                     &      %
                               & 1.77                & 1.77 %
                               & 1.77                & 1.77 \\%
$d_{C-Si}$~(\AA)$^b$           &                     &      %
                               &                     &      %
                               & 1.94                & 1.94 %
                               & 1.95                & 1.95 \\%
$d_{C-C}$~(\AA)$^b$            &                     &      %
                               &                     &      %
                               & 1.36                & 1.36 %
                               & 1.42                & 1.42 \\%
\hline %
$E_{coh}$~(eV/atom)$^c$        & 4.74                & 4.74    %
                               & 4.98                & 4.98    %
                               & 5.25                & 5.26    %
                               & 5.54                & 5.54  \\%
\modR{ %
$E_{form}$~(eV/atom)$^d$
}%
& %
\modR{ %
-0.01
}%
& %
\modR{ %
-0.01
}%
& %
\modR{ %
-0.25
}%
& %
\modR{ %
-0.24
}%
& %
\modR{ %
0.68
}%
& %
\modR{ %
0.68
}%
& %
\modR{ %
0.39
}%
& %
\modR{ %
0.39
}%
  \\%
$c_{11}$~(N/m)$^e$             & 7.17                & 40.59   %
                               & 10.24               & 57.53   %
                               & 6.42                & 33.79   %
                               & 14.89               & 53.35 \\%
$c_{22}$~(N/m)$^e$             & 156.02              & 144.13  %
                               & 189.59              & 173.68  %
                               & 145.94              & 124.02  %
                               & 199.72              & 185.35\\%
\hline %
$E_{g-PBE}$~(eV)$^f$           & 0.52                & 0.80    %
                               & 1.39                & 1.26    %
                               & 0.71                & 0.70    %
                               & 0.00                & 0.00  \\%
$E_{g-HSE06}$~(eV)$^g$         & 1.38                & 1.69    %
                               & 2.15                & 2.06    %
                               & 1.57                & 1.40    %
                               & 0.00                & 0.00  \\%
\hline\hline
\end{tabular}
\label{table1}
\begin{tablenotes}
\item[1] %
$^a$ $|\vec{a_1}|$ and $|\vec{a_2}|$ are the in-plane lattice
     constants defined in Fig.~\protect\ref{fig1}. \\%
$^b$ $d$ is the equilibrium bond length between the respective
     species. \\%
$^c$ $E_{coh}$ is the cohesive energy per average atom with
     respect to isolated atoms. \\%
\modR{ %
$^d$ $E_{form}$ is the formation energy per average atom with
     respect to elemental structures.
}%
\\%
$^e$ $c_{11}$ ($c_{22}$) are the 2D elastic constants
     along the $x$ ($y$) direction. \\%
$^f$ $E_{g}$-PBE   is the band gap value at the DFT-PBE level. \\%
$^g$ $E_{g}$-HSE06 is the band gap value at the DFT-HSE06 level. \\%
$^h$ Length of bonds within the structural plane. \\%
$^i$ Length of bonds out of the structural plane.
\end{tablenotes}
\end{table*}

\begin{figure}[b]
\includegraphics[width=1.0\columnwidth]{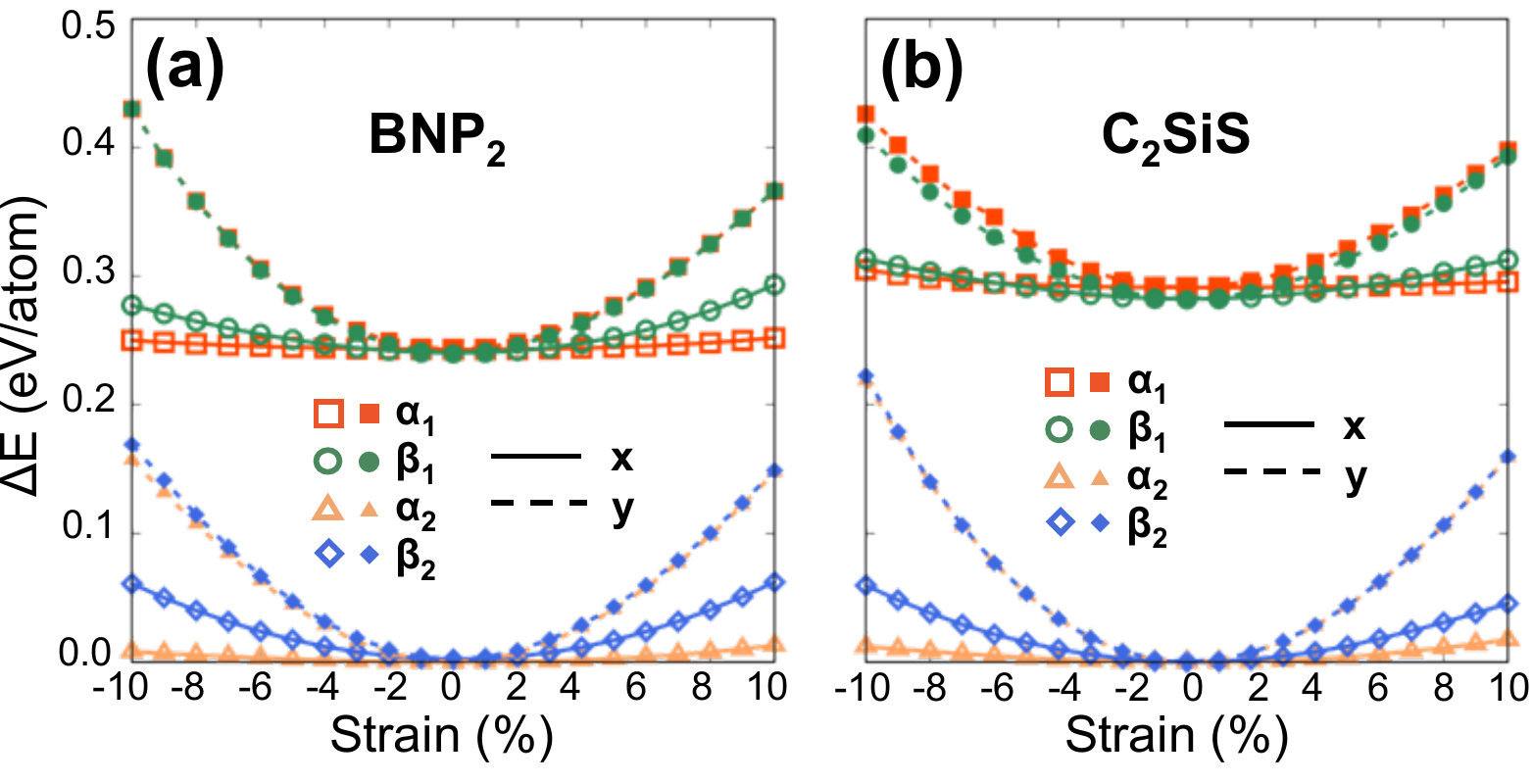}
\caption{Effect of uniaxial in-layer strain on the relative
binding energy ${\Delta}E$ in monolayers of %
(a) BNP$_2$ and %
(b) C$_2$SiS. Results for different allotropes are distinguished
by color and symbols. Results for strain along the $x$-direction
are shown by solid lines and for strain along the $y$-direction by
dashed lines.
\label{fig2}}
\end{figure}

\begin{figure*}[t]
\includegraphics[width=1.8\columnwidth]{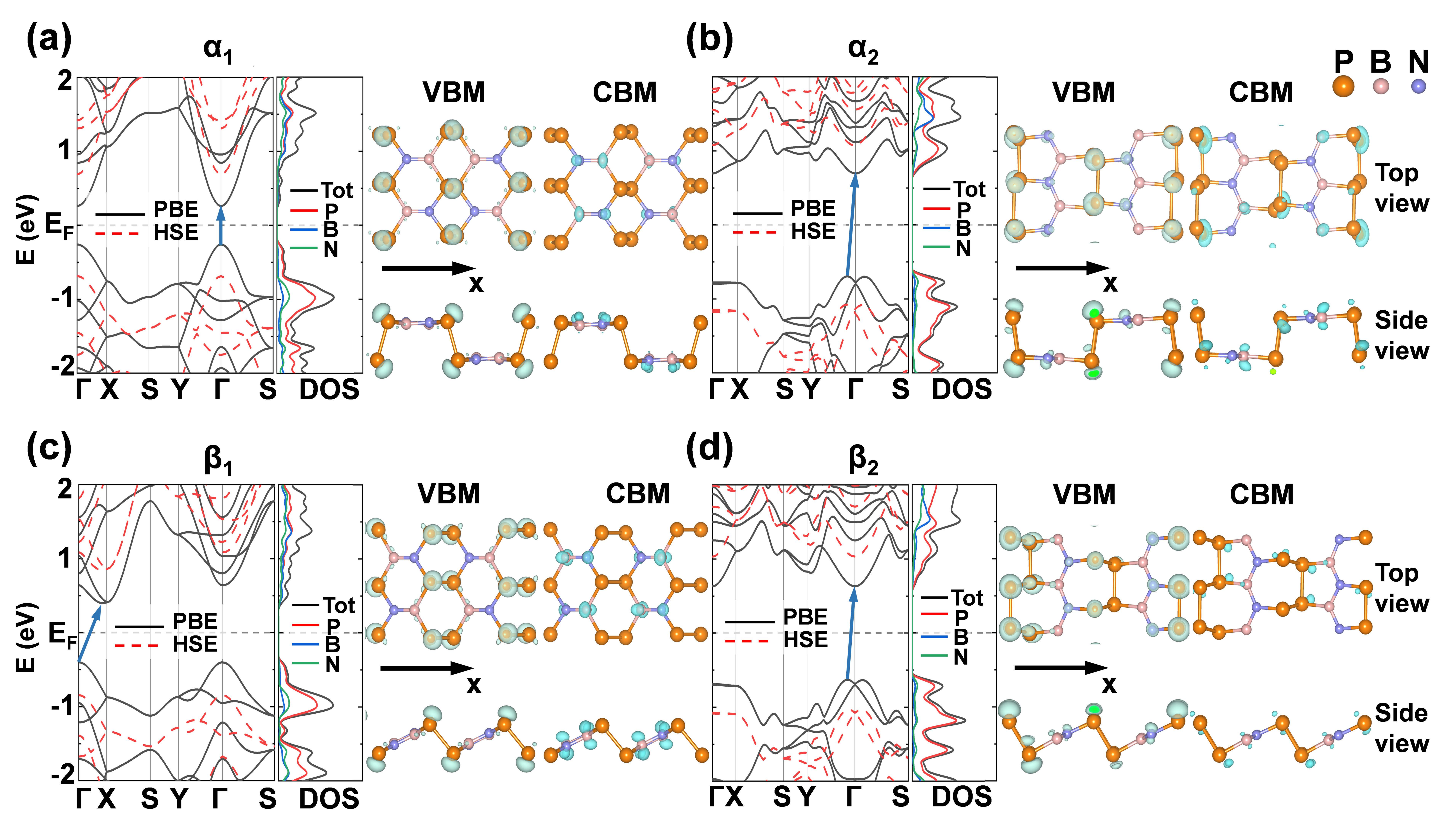}
\caption{ %
Electronic structure of a BNP$_2$ monolayer. %
Shown are the calculated band structure $E(${\bf{k}}) and %
the density of states (DOS) including its projection on
individual atoms (left panels), %
the partial charge density distributions at the %
valence band maximum (VBM) %
and the conduction band minimum (CBM) (right panels) of the %
(a) $\alpha_1$, %
(b) $\alpha_2$, %
(c) $\beta_1$, and %
(d) $\beta_2$ %
allotropes. %
All results are based on the DFT-PBE functional with the exception
of DFT-HSE06 data, which are shown by the red dashed lines in the
band structure plots. The blue arrows indicate the fundamental
band gap. Contributions of different atoms are distinguished by
color in the PDOS plots. Isosurface plots of the partial charge
density are presented at the isosurface value of
$8.0{\times}10^{-3}$~e/Bohr$^3$ for all allotropes. %
\label{fig3}}
\end{figure*}

\section*{Computational Techniques}

We have used {\em ab initio} density functional theory (DFT), as
implemented in the \textsc{VASP}
code~\cite{{VASP},{VASP2},{VASP3}}, throughout the study. We
applied periodic boundary conditions, with 2D structures separated
by a vacuum region in excess of $20$~{\AA}. The reciprocal space
was sampled by a fine grid~\cite{Monkhorst-Pack76} of
$8{\times}6{\times}1$~$k$-points in the Brillouin zone of 8-atom
unit cells or its equivalent in supercells. We used
projector-augmented-wave (PAW) pseudopotentials~\cite{PAWPseudo}
and the Perdew-Burke-Ernzerhof (PBE)~\cite{PBE}
exchange-correlation functionals. Selective band structure
calculations were performed using the hybrid HSE06
functional~\cite{{HSE06-1},{HSE06-2}} with the default mixing
parameter ${\alpha}=0.25$. We utilized the DFT-D2{~\cite{DFTD2}}
method to represent van der Waals (vdW) corrections to the total
energy. We used a cutoff energy of 500~eV for the plane-wave basis
set and considered the electronic structure to be converged once
the total energy difference between subsequent electronic
structure iterations would not exceed $10^{-5}$~eV. All geometries
were optimized using the conjugate-gradient
method~\cite{CGmethod}, until none of the residual
Hellmann-Feynman forces exceeded $10^{-2}$~eV/{\AA}. The phonon
calculations were carried out using the density functional
perturbation theory~\cite{{DFPT01},{DFPT02},{DFPT03}}, as
implemented in the \textsc{PHONOPY} code~\cite{phonopy}. The
system dynamics was studied using canonical {\em ab initio} MD
simulations with $3$~fs time steps. Our results are for supercells
containing more than 100 atoms, which were kept at
\modR{ %
temperatures
}%
of $300$~K, $500$~K or $1000$~K for periods exceeding $15$~ps.

\section*{Results and Discussion}

\subsection*{Structure of BNP$_2$ and C$_2$SiS 2D allotropes}

As introduced above, the 2D compounds BNP$_2$ and C$_2$SiS are
isoelectronic to previously reported PC monolayer
structures~\cite{2DPC16}. We have identified four 2D allotropes,
called $\alpha_1$, $\beta_1$, $\alpha_2$ and $\beta_2$, for each
of the systems. The most stable structures of each allotrope are
shown in Fig.~\ref{fig1}. All atoms are 3-fold coordinated,
causing a coexistence of $sp^2$ and $sp^3$ bonding and leading to
structural ridges in the geometry of all allotropes. When viewed
from the side, the $\alpha$ allotropes have an armchair profile,
whereas the $\beta$ allotropes have a zigzag profile. With the
exception of $\beta_1$-C$_2$SiS with 8 atoms in the unit cell, all
allotropes have rectangular unit cells containing 16 atoms.
Additional metastable structures with different atomic
arrangements in the unit cell are discussed in the Appendix.

As seen in Fig.~\ref{fig1}, the optimized $\alpha_1$ and $\beta_1$
allotropes of BNP$_2$ structures consist of isolated P-P and B-N
dimers forming a 2D hexagonal structure. Since the orientation of
the B-N dimers alternates in the plane of the system, there is no
net dipole moment in the system.
\modR{ %
P atoms %
}%
prefer the $sp^3$
configuration with a lone electron pair, whereas B and N atoms
prefer the $sp^2$ configuration in all BNP$_2$ allotropes. The
$\alpha_1$ and $\beta_1$ allotropes represent two equivalent ways
to achieve
\modR{ %
optimum configuration %
}%
with the same topology, in analogy to black and blue
phosphorene.~\cite{DT230}

The $\alpha_2$ and $\beta_2$ allotropes of BNP$_2$ contain
alternating chains of P and
\modR{ %
BN
}%
along the $y$-direction. The
mismatch between the equilibrium P-P and B-N bond lengths leads to
the formation of pentagon-heptagon pairs instead of hexagons.

The structure of all 2D allotropes of C$_2$SiS closely resembles
that of BNP$_2$ due to the coexistence of $sp^2$-bonded C atoms
and $sp^3$-bonded Si and S atoms. Similar to BNP$_2$, the
$\alpha_1$ and $\beta_1$ allotropes of C$_2$SiS contain isolated
Si-S and C-C dimers. Also in this case, there are two
distinguishable, topologically equivalent geometries. Even though
the polar Si-S bonds are out of plane in this compound, their
alternating orientation eliminates a net dipole moment. The
$\alpha_2$ and $\beta_2$ allotropes of C$_2$SiS, analogous to
those of BNP$_2$, contain alternating C and SiS chains along the
$y$-direction. Bond length mismatch leads to a preferential
formation of pentagon-heptagon pairs in the layer.

A summary of the structural characteristics and the cohesive
energy of all these allotropes is presented in Table~\ref{table1}.
In the BNP$_2$ system, the B-N dimers in $\alpha_1$ and $\beta_1$
allotropes are connected by typical double bonds with
$d_{BN}{\approx}1.40$~{\AA}, close to the $1.403$~{\AA} value in
amino borane~\cite{SUGIE79}. The bond length of the B-N chains in
$\alpha_2$ and $\beta_2$ allotropes is slightly longer, at about
$1.46$~{\AA}, close to the $1.45$~{\AA} value in 2D
$h$-BN~\cite{BN2009}. These results are consistent with the $sp^2$
configuration of B and N atoms in 2D BNP$_2$.

\begin{figure}
\includegraphics[width=0.8\columnwidth]{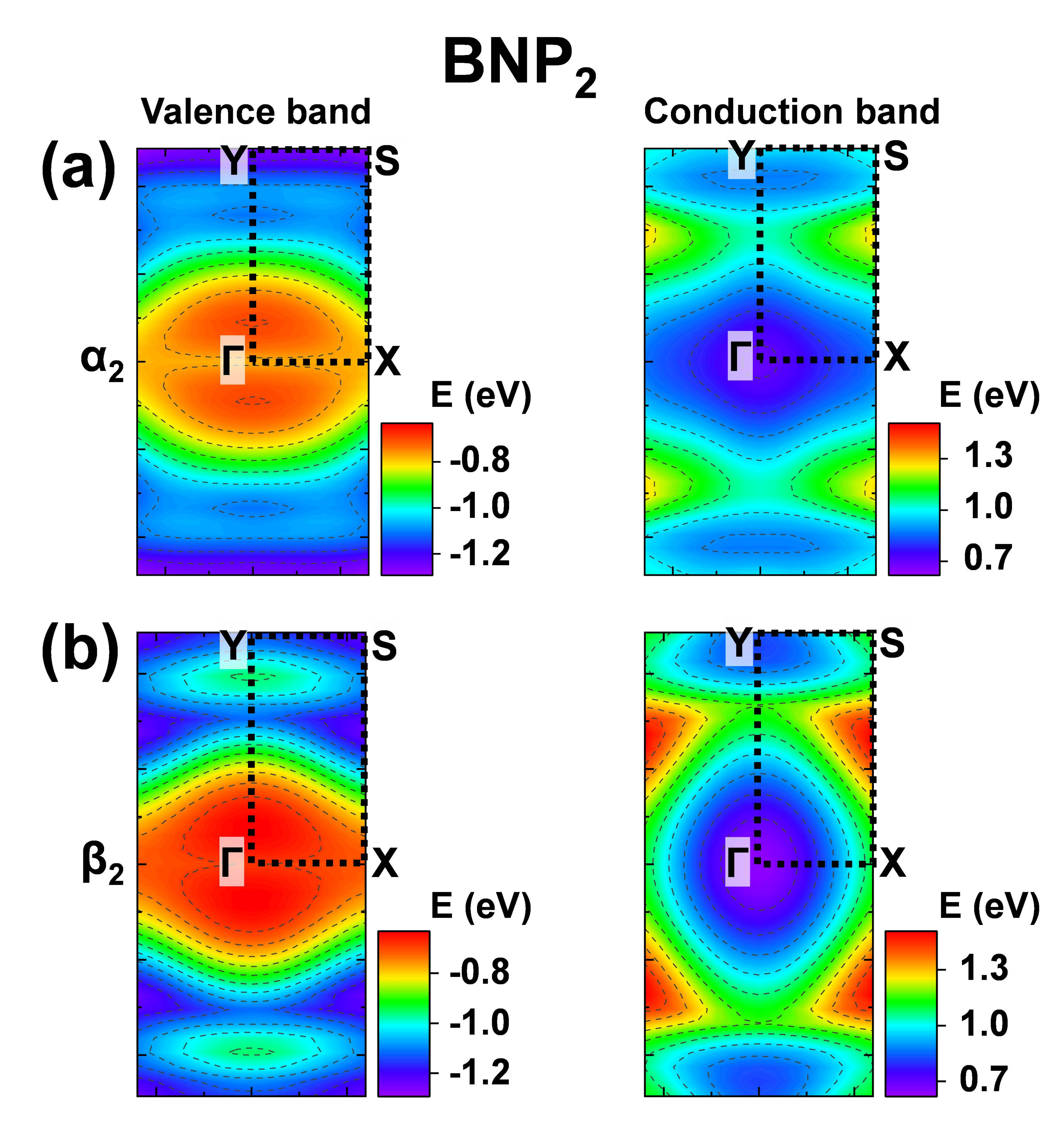}
\caption{%
Contour plots of the energy dispersion $E$({\bf{k}}) of the top
valence band (left panels) and the lowest conduction band (right
panels) in %
(a) $\alpha_2$-BNP$_2$ and %
(b) $\beta_2$-BNP$_2$. %
The Fermi level is set as the reference energy of zero. %
\label{fig4}}
\end{figure}

\begin{figure*}[t]
\includegraphics[width=1.8\columnwidth]{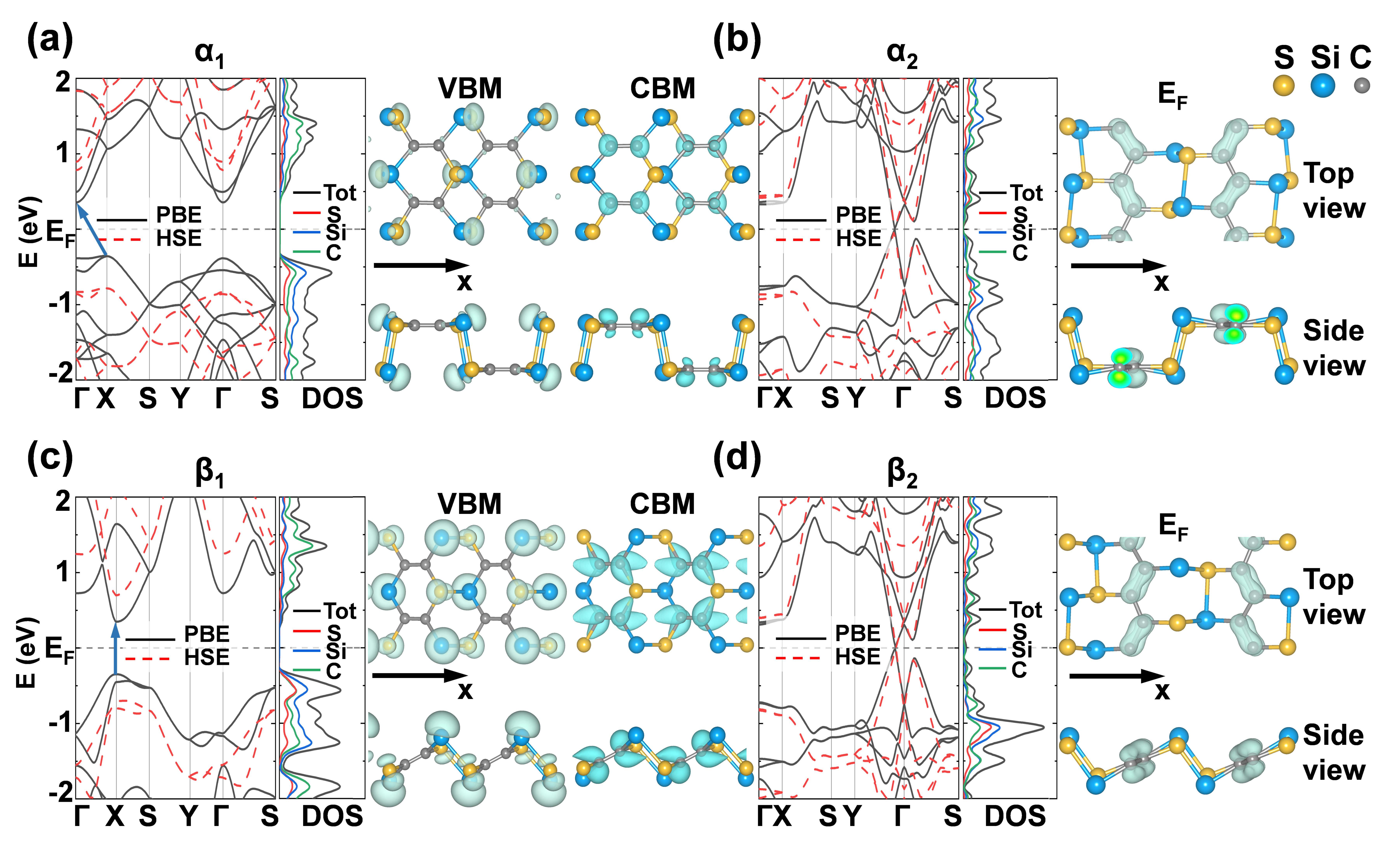}
\caption{%
Electronic structure of a C$_2$SiS monolayer. %
The calculated band structure $E(${\bf{k}}) and %
the density of states (DOS) including its projection on
individual atoms are shown in the left panels. %
The partial charge density distributions at the %
valence band maximum (VBM) %
and the conduction band minimum (CBM) %
for the semiconducting allotropes and for the frontier states in
the semimetallic allotropes are shown in the right panels. %
Results
\modR{ %
are
}%
presented for the %
(a) $\alpha_1$, %
(b) $\alpha_2$, %
(c) $\beta_1$, and %
(d) $\beta_2$ allotropes. %
All results are based on the DFT-PBE functional with the exception
of DFT-HSE06 data, which are shown by the red dashed lines in the
band structure plots. The blue arrows indicate the fundamental
band gap in the semiconducting allotropes. Contributions of
different atoms are distinguished by color in the PDOS plots.
Isosurface plots of the partial charge density are presented at
the isosurface value of
$8.0{\times}10^{-3}$~e/Bohr$^3$ for all allotropes. %
\label{fig5}}%
\end{figure*}

\begin{figure}[b]
\includegraphics[width=1.0\columnwidth]{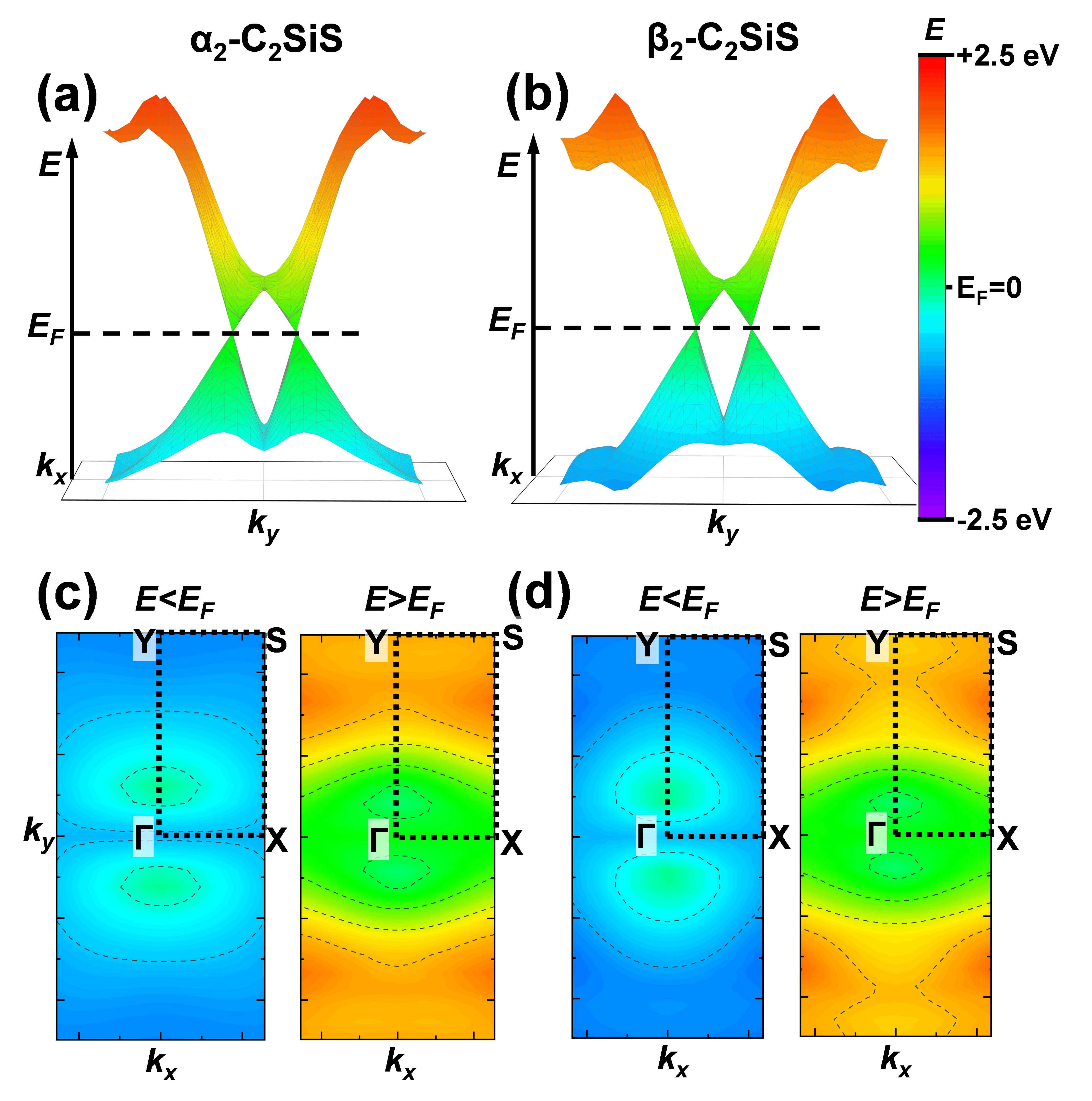}
\caption{%
Band dispersion in the semimetallic allotropes %
$\alpha_2$-C$_2$SiS (left) and $\beta_2$-C$_2$SiS (right) near the
Fermi level. $E(${\bf{k}}) results are presented as 3D plots in
(a),(b) and 2D contour plots in (c),(d). The energy with respect
to $E_F=0$ is represented by color.
\label{fig6}}%
\end{figure}

We find similar structural characteristics in the C$_2$SiS system.
As expected for $sp^2$-bonded C atoms, the C-C bond lengths are
comparable to the double bonds in ethylene %
and the conjugated bonds in benzene %
or graphene. %
The length of all the other single bonds appears rather insensitive to
the structure.

It is also noteworthy that P-N bonds are slightly shorter than P-B
bonds in BNP$_2$ due to the smaller atomic radius of the more
electronegative element. The same is true, for the same reason,
with the C-S bonds that are slightly shorter than C-Si bonds in
C$_2$SiS allotropes.

For the sake of easy comparison, we define the cohesive energy
$E_{coh}$ of an ``average'' atom by dividing the total atomization
energy by the total number of atoms. Our results for $E_{coh}$ are
listed in Table~\ref{table1}. We found that for both BNP$_2$ and
C$_2$SiS, $\alpha$ and $\beta$ allotropes with the same index are
almost equally stable. Our results also indicate that for both
BNP$_2$ and C$_2$SiS, $\alpha_2$ and $\beta_2$ allotropes are
energetically more stable than $\alpha_1$ and $\beta_1$ allotropes
by ${\Delta}E_{coh}{\approx}0.2-0.3$~eV/atom. We find the C$_2$SiS
structures to be generally more stable than BNP$_2$ structures.
The listed values of $E_{coh}{\approx}5$~eV/atom here are
comparable to cohesive energies found in PC
monolayers~\cite{2DPC16}.

We have further confirmed the stability of each allotrope by
calculating the phonon spectra and by performing {\em ab initio} MD
simulations.
Details of all these calculations are presented in the Appendix.

\modR{ %
We also calculated the formation energy per `average atom'
$E_{form}$ to investigate the relative stability of the ternary
allotropes comparing with respect to their elemental components.
We defined $E_{form}$ as
\begin{eqnarray}
E_{form}(X_lY_mZ_n) &=& E(X_lY_mZ_n) \nonumber \\
&-& \frac{l{\cdot}E(X) + m{\cdot}E(Y) + n{\cdot}E(Z)}{l+m+n} , %
\label{eq1}
\end{eqnarray}
where $E(X_lY_mZ_n)$, $E(X)$, $E(Y)$, $E(Z)$ are the respective
total energies per average atom of the X$_l$Y$_mZ_n$ compound and
the elemental structures of X, Y and Z. In the particular case of
2D BNP$_2$ and C$_2$SiS structures, the elemental structures we
consider are bulk boron, gas of N$_2$ molecules, phosphorene,
graphene, bulk silicon and bulk sulfur. The results of $E_{form}$
for all the allotropes are listed in Table~\ref{table1}.
$E_{form}>0$ indicates instability and possibility of exothermic
decomposition, whereas $E_{form}<0$ indicates stability. %
}%

\modR{ %
Our results for $E_{form}$ in Table~\ref{table1} indicate that all
BNP$_2$ allotropes are stable and will not decompose below
${\approx}1000$~K, at which point bonds may be broken. The values
of $E_{form}$ for C$_2$SiS allotropes considered here are
positive, but rather small. The values $E_{form}=0.68$~eV/atom for
the $\alpha_1$ and $\beta_1$ phases and $E_{form}=0.39$~eV/atom
for the $\alpha_2$ and $\beta_2$ phases are an order of magnitude
smaller than cohesive energies and comparable to
$E_{form}=0.54$~eV/atom found in 2D-PC~\cite{2DPC16}. Structural
changes would require overcoming a significant activation barrier
associated with breaking bonds and protect C$_2$SiS allotropes
from structural changes at room temperature. As mentioned above,
interatomic bonds may break and structural changes may become
possible at extremely high temperatures of ${\approx}1000$~K. We
find these estimates confirmed by our MD simulation results
presented in the Appendix.
}%

\subsection*{Anisotropic in-plane stiffness of 2D BNP$_2$ and
C$_2$SiS allotropes}

To investigate the elastic response of the 2D ternary structures,
we subjected all allotropes in this study to in-layer uniaxial
strain along the $x$- and $y$-direction and display differences
${\Delta}E_{coh}$ of the average binding energy with respect to
the most stable allotrope in Fig.~\ref{fig2}. The presence of
structural ridges renders all allotropes much softer along the
$x$-direction normal to the
\modR{ %
ridge direction.
}%
The
$\alpha$ allotropes of both BNP$_2$ and C$_2$SiS are particularly
soft along the $x$-direction, with
${\Delta}E{\lesssim}$18~meV/atom when
\modR{ %
subjected
}%
to ${\le}10$\%
compressive or tensile strain. The 2D elastic moduli~\cite{DT255}
$c_{ii}$ along the $x$- and $y$-direction, obtained using these
calculations, are summarized in Table~\ref{table1}. For the
semiconducting allotropes, these values play an important role in
carrier mobilities.


\subsection*{Electronic structure of BNP$_2$ 2D allotropes}

Results of our DFT calculations for the electronic structure of
BNP$_2$ monolayers are presented in Fig.~\ref{fig3}. Our DFT-PBE
results show all the four allotropes to be semiconductors. The
fundamental band gap is direct in $\alpha_1$-BNP$_2$ and indirect
in all the other BNP$_2$ allotropes. Band gap values $E_g$ based
on DFT-PBE Kohn-Sham energies, ranging from $0.52$~eV to
$1.39$~eV, are listed in Table~\ref{table1}. We should note here
that the interpretation of Kohn-Sham eigenvalues as self-energies
is strictly incorrect and that DFT-based band gaps are typically
underestimated. We find this to be the case when comparing band
gap values based on DFT-PBE and the hybrid DFT-HSE06 functional in
Table~\ref{table1}. As seen when comparing band structure results
based on DFT-PBE, given by the solid lines in Fig.~\ref{fig3}, and
DFT-HSE06, shown by the red dashed lines, the main difference is
the opening of the fundamental band gap in DFT-HSE06, whereas the
band dispersion is unaffected. We also present the total
electronic density of states (DOS) and its projection onto the
different species next to the band structure plots.

Contour plots of the projected DOS (PDOS) of all BNP$_2$
allotropes at the valence band maximum (VBM) and conduction band
minimum (CBM) are shown in Fig.~\ref{fig3}, with the contributions
\modR{ %
to VBM and CBM
}%
distinguished by color. We find the character
of the VBM to be dominated by the lone-pair states of the $sp^3$
hybridized P atoms in all BNP$_2$ allotropes. The character of the
CBM is very different. The dominant contribution in $\alpha_1$ and
$\beta_1$ allotropes of BNP$_2$ comes from out-of-plane $p$
orbitals of $sp^2$ hybridized B and N atoms. In the $\alpha_2$ and
$\beta_2$ allotropes, $p$ orbitals of all the atoms contribute to
the CBM character.

We also observe an unusually flat band in the band structure of
$\alpha_2$ and $\beta_2$ allotropes of BNP$_2$ at the top of the
valence region, along the ${\Gamma}-X$ line.
To further explore the origin of this flat band, we show 2D contour
plots of the dispersion within the top valence and bottom conduction
bands in Fig.~\ref{fig4}.

In both allotropes, the CBM energy minimum occurs at the $\Gamma$
point and the band dispersion around is almost isotropic and
free-electron like. In contrast, the VBM is displaced from the
$\Gamma$ point in the direction towards $X$ for both allotropes.
Whereas the band dispersion is almost flat along the
$x$-direction, it is significantly higher along $y$-direction near
the VBM, indicating a significant anisotropy in the carrier
effective mass. As seen in Fig.~\ref{fig3}(b) and (d), this is
caused at the VBM by a larger separation and lower hybridization
between the P lone pair states along the $x$-direction than along
the $y$-direction.

\subsection*{Electronic structure of C$_2$SiS 2D allotropes}

Results of our DFT calculations for the electronic structure of
C$_2$SiS monolayers are presented in Fig.~\ref{fig5}. Our findings
indicate that only the $\alpha_1$ and the $\beta_1$ allotropes are
semiconducting, whereas the $\alpha_2$ and $\beta_2$ allotropes
are semimetallic. The band gap is indirect in $\alpha_1$-C$_2$SiS
and direct in $\beta_1$-C$_2$SiS. Numerical band gap values based
on DFT-PBE and DFT-HSE06 are listed in Table~\ref{table1} and fall
in the range of values found in BNP$_2$.

According to Fig.~\ref{fig5}, the character of VBM and CBM states
in the semiconducting allotropes $\alpha_1$ and $\beta_1$ is very
similar. In analogy to results for BNP$_2$, the VBM of C$_2$SiS is
dominated by lone-pair orbitals of Si atoms and the CBM contains
mostly out-of-local-plane $p$ orbitals of C atoms.

The $\alpha_2$ and $\beta_2$ allotropes of C$_2$SiS are found to
be semimetallic by both DFT-PBE and DFT-HSE06. Dirac ``cones'' are
seen in the band structure in Fig.~\ref{fig5}(b) and (d), with the
Dirac point at $E_F$ located between $\Gamma$ and $Y$ in the
Brillouin zone. The frontier states at $E_F$ are dominated by
out-of-plane $p$ orbitals on C atoms.

The Dirac ``cones'' of $\alpha_2$- and $\beta_2$-C$_2$SiS are
visualized in more detail in Fig.~\ref{fig6}. The band dispersion
in both allotropes is linear around $E_F$, but anisotropic. As
seen in Fig.~\ref{fig6}(c) and (d), the cross-section of the
``cone'' at $E<E_F$ or $E>E_F$ is not a circle, but rather an
ellipse elongated in the $k_x$-direction. The higher steepness of
the band dispersion along the $y$-direction near the Dirac point
is a consequence of a stronger interaction between C$2p$ orbitals
within zigzag chains of carbons, aligned with the $y$-direction.
These C chains in $\alpha_2$-C$_2$SiS and $\beta_2$-C$_2$SiS are
clearly visible in Figs.~\ref{fig1}(g) and (h). Otherwise, these
``cones'' are very similar in Dirac point location, band
dispersion and its anisotropy in both allotropes.

\begin{table*}
\caption{%
Calculated carrier mobilities and related quantities in
semiconducting BNP$_2$ and C$_2$SiS allotropes. %
}%
\setlength{\tabcolsep}{3mm}
\begin{tabular}{lcccccccccc}
\hline\hline
                     &                     & $m_x^*$$^a$  %
                     & $m_y^*$$^a$  %
                     & $E_{1x}$$^b$        & $E_{1y}$$^b$ %
                     & $\mu_x$$^c$         & $\mu_y$$^c$  \\%
          &          &\multicolumn{2}{c}{($m_0$)} &\multicolumn{2}{c}{(eV)} %
                     &\multicolumn{2}{c}{(10$^3$ cm$^2$V$^{-1}$s$^{-1}$)} \\%
\hline %
\multirow{2}{*}{$\alpha_1$-BNP$_2$}
                         & $e$                 & 0.51 %
                         & 0.34 %
                         & 1.36                & 0.40 %
                         & 0.40                & 151.21 \\%
                         & $h$                 & 0.75 %
                         & 0.39 %
                         & 1.83                & 1.60 %
                         & 0.11                & 6.16 \\%
\multirow{2}{*}{$\beta_1$-BNP$_2$}
                         & $e$                 & 0.67 %
                         & 0.34 %
                         & 0.69                & 1.37 %
                         & 5.89                & 10.51 \\%
                         & $h$                 & 0.67 %
                         & 0.57 %
                         & 0.37                & 1.26 %
                         & 15.82               & 5.62 \\%
\multirow{2}{*}{$\alpha_2$-BNP$_2$}
                         & $e$                 & 0.88 %
                         & 0.78 %
                         & 1.01                & 0.88 %
                         & 0.30                & 8.29 \\%
                         & $h$                 & 3.13 %
                         & 0.50 %
                         & 0.10                & 5.07 %
                         & 5.67                & 0.26 \\%
\multirow{2}{*}{$\beta_2$-BNP$_2$}
                         & $e$                 & 0.53 %
                         & 0.75 %
                         & 2.41                & 0.68 %
                         & 0.65                & 17.17 \\%
                         & $h$                 & 1.63 %
                         & 0.52 %
                         & 0.90                & 5.96 %
                         & 1.03                & 0.22 \\%
\multirow{2}{*}{$\alpha_1$-C$_2$SiS}
                         & $e$                 & 0.53 %
                         & 0.28 %
                         & 0.39                & 1.37 %
                         & 4.38                & 15.66 \\%
                         & $h$                 & 1.72 %
                         & 0.76 %
                         & 2.54                & 0.87 %
                         & 0.01                & 4.85 \\%
\multirow{2}{*}{$\beta_1$-C$_2$SiS}
                         & $e$                 & 0.14 %
                         & 1.57 %
                         & 2.51                & 6.31 %
                         & 1.69                & 0.09 \\%
                         & $h$                 & 0.78 %
                         & 2.24 %
                         & 1.87                & 1.42 %
                         & 0.20                & 0.45 \\%
\hline\hline
\end{tabular}
\label{table2}
\begin{tablenotes}
\item[1] %
Electrons are denoted by $e$ and holes by $h$. \\%
$^a$ $m_x^*$ ($m_y^*$) are the carrier effective masses
     along the $x$ ($y$) direction. \\%
$^b$ $E_{1x}$ ($E_{1y}$) are the deformation potentials
     along the $x$ ($y$) direction. \\%
$^c$ $\mu_x$ ($\mu_y$) are the carrier mobilities
     along the $x$ ($y$) direction at $300$~K.
\end{tablenotes}
\end{table*}

\subsection*{Carrier mobilities in semiconducting BNP$_2$ and
             C$_2$SiS 2D allotropes}

Anisotropy in semiconducting BNP$_2$ and C$_2$SiS 2D allotropes is not
only limited to the geometry, the elastic response and the electronic
structure, but is also present in the carrier mobility. In absence of
defects and external scattering centers, the mobility of carriers in 2D
semiconductors is limited by acoustic phonons. We calculated the
carrier mobilities of BNP$_2$ and C$_2$SiS 2D systems along the $x$-
and $y$-direction using the deformation potential theory
expression~\cite{{DPtheory},{BPmob},{Wang16PC}}
\begin{equation}
{\mu}_{i} = \frac{e{\hbar}^3c_{ii}}{k_BTm_i^*m_d{E_{1i}}^2} \;. %
\label{eq2}
\end{equation}
Here, $i$ represents the Cartesian direction, with $i=1$ standing
for $x$ and $i=2$ for $y$, and $e$ is the carrier charge. $c_{ii}$
is the 2D elastic modulus~\cite{DT255} along the direction $i$,
obtained from the strain energy curve in Fig.~\ref{fig2}. $T$ is
the temperature, $m_i^*$ is the effective mass along the
$i$-direction, and $m_d$ is the average effective mass given by
$m_d=({m_x^*}{m_y^*})^{1/2}$. The deformation potential $E_{1i}$
along the $i$-direction is determined at the valence band maximum
(VBM) for holes and the conduction band minimum (CBM) for
electrons. It is defined by $E_{1i} =
{\Delta}V/({\Delta}a_i/a_i)$, where ${\Delta}V$ is the energy
shift of the band edge with respect to the vacuum level under a
small change ${\Delta}a_i$ of the lattice constant $a_i$. Our
results for room-temperature carrier mobilities in all
semiconducting 2D allotropes are summarized in Table~\ref{table2}.

We found that most allotropes in our study exhibit very high and
remarkably anisotropic carrier mobilities. The highest carrier
mobility value we found is
${\mu_y}=1.51{\times}10^5$~cm$^2$V$^{-1}$s$^{-1}$ for electrons in
$\alpha_1$-BNP$_2$ along the $y$-direction, over two orders of
magnitude higher than the value
${\mu_x}=0.4{\times}10^3$~cm$^2$V$^{-1}$s$^{-1}$ along the
$x$-direction. An important contributing factor to the mobility is
the 2D elastic modulus~\cite{DT255}. The superior electron
mobility along its $y$-direction is a result of increased
stiffness along this direction. According to Table~\ref{table1},
$c_{22}$ is indeed much larger than $c_{11}$ for most allotropes
we study.

Unlike for electrons, hole mobility is highest along the $x$-direction
in $\beta_1$, $\alpha_2$ and $\beta_2$ allotropes of BNP$_2$. The
change of preferential transport direction is caused by substantial
deformation potential anisotropy $E_{1x}<E_{1y}$ at the VBM according
to Table~\ref{table2}. Carrier mobilities in semiconducting 2D
allotropes of C$_2$SiS are comparable to those in BNP$_2$.

Carrier mobilities %
${\mu}>10^5$~cm$^2$V$^{-1}$s$^{-1}$ %
in $\alpha_1$-BNP$_2$ with
\modR{ %
$E_g>0.5$~eV
}%
and %
${\mu}{\gtrsim}10^4$~cm$^2$V$^{-1}$s$^{-1}$ %
in $\alpha_2$ and $\beta_2$ allotropes of BNP$_2$ with
\modR{ %
$E_g>1$~eV, evaluated at the PBE level of DFT
}%
are unusually high in view of the wide band gaps.
\modR{ %
This is particularly evident when compared to the values found in
other similar 2D materials, as discussed in the Appendix.
}%
Even though these mobilities will be reduced due to inevitable
defects and the interaction with the substrate under realistic
conditions, we expect values in excess of
${\mu}=0.3{\times}10^3$~cm$^2$V$^{-1}$s$^{-1}$, which has been
reported for phosphorene~\cite{DT229}. These results are very
promising for the realization of 2D semiconducting devices with
high ON/OFF ratios and on-state currents.

\section*{Summary and Conclusions}

In summary, we have introduced previously unexplored 2D ternary
compounds BNP$_2$ and C$_2$SiS as isoelectronic counterparts of 2D
PC structures. Using {\em ab initio} density functional theory, we
have identified four stable allotropes of each compound and
confirmed their stability by calculated phonon spectra and
molecular dynamics simulations. All allotropes display structural
and elastic anisotropy due to structural ridges in their geometry,
which are caused by coexisting $sp^2$ and $sp^3$ hybridization.
Whereas all BNP$_2$ allotropes are semiconducting, we find only
two allotropes of C$_2$SiS to be semiconducting. The other two
allotropes of C$_2$SiS are semimetallic and show anisotropic Dirac
cones at $E_F$. The fundamental band gaps of the semiconducting
allotropes we study range from
\modR{ %
$0.5$~eV to $1.4$~eV at the PBE level and $1.4$~eV to $2.2$~eV at
the HSE06 level of DFT. These 2D systems %
}%
display carrier mobilities as high as
$1.5{\times}10^5$~cm$^2$V$^{-1}$s$^{-1}$. Such high mobilities,
with two orders of magnitude in anisotropy ratio, are desirable,
but quite uncommon in semiconductors with so wide band gaps.
Combination of wide band gaps with high and anisotropic carrier
mobilities offer great promise for applications of 2D BNP$_2$ and
C$_2$SiS structures in electronics and optoelectronics.

\begin{acknowledgments}
This study was supported by the National Natural Science Foundation of
China (NSFC) under Grant No. 61704110, the Fundamental Research Fund
for the Central Universities, the Shuangchuang Doctoral Program of the
Jiangsu Province, and by the Zhongying Young Scholar Program of
Southeast University. D.T. acknowledges financial support by the
NSF/AFOSR EFRI 2-DARE grant number EFMA-1433459. We thank the Big Data
Computing Center of Southeast University for providing facility support
for performing calculations presented in this manuscript.
\end{acknowledgments}

\section*{Appendix}
\setcounter{figure}{0}
\renewcommand\thefigure{A\arabic{figure}}
\setcounter{table}{0}
\renewcommand\thetable{A\arabic{table}}
\setcounter{equation}{0}
\renewcommand{\theequation}{A\arabic{equation}}


\subsection*{Metastable structures of 2D BNP$_2$ and C$_2$S\lowercase{i}S}

\begin{figure}
\includegraphics[width=1.0\columnwidth]{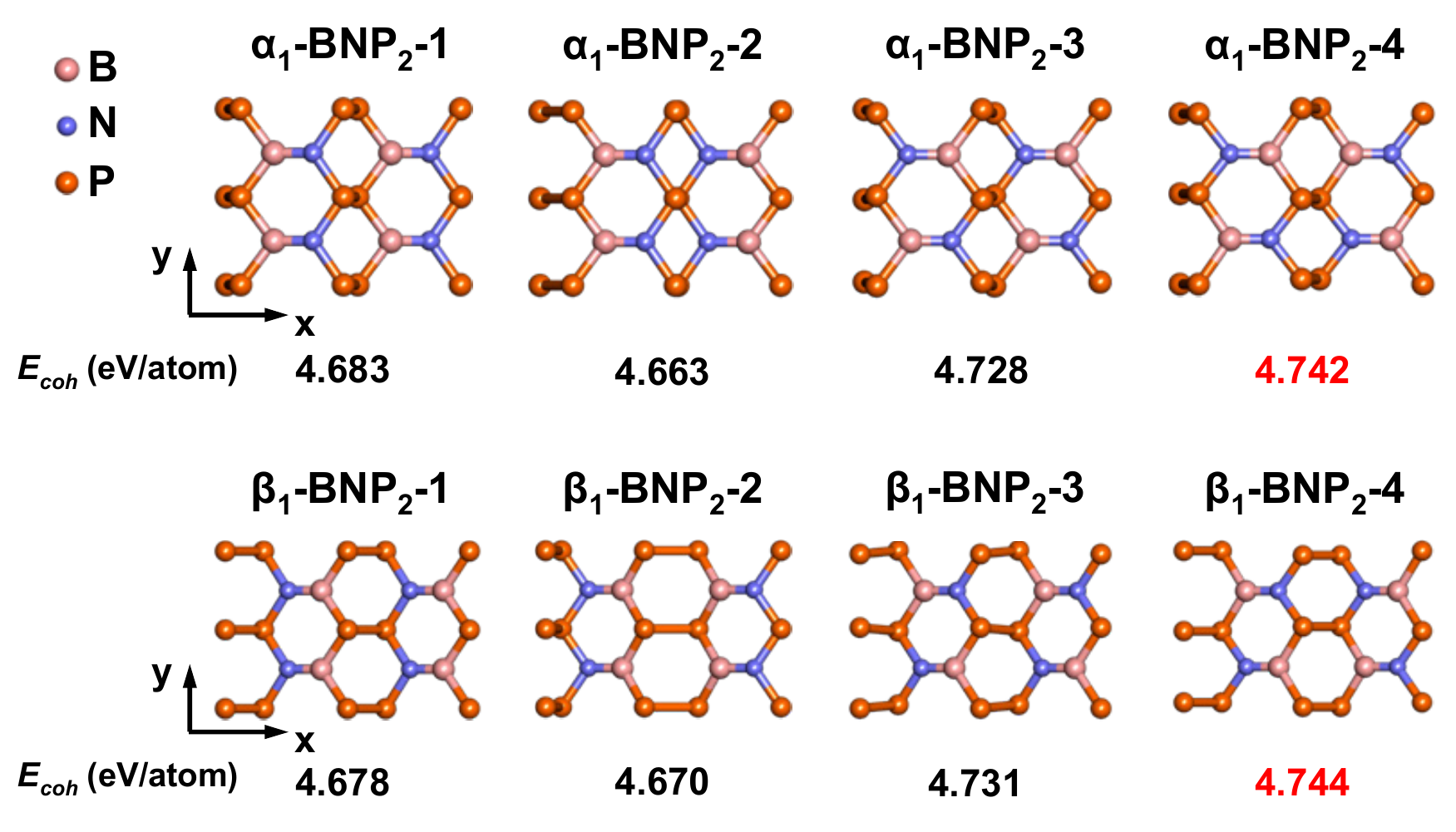}
\caption{Ball-and-stick model of metastable structures related to
the $\alpha_1$ and $\beta_1$ allotropes of BNP$_2$, shown in top
view. Cohesive energies $E_{coh}$ are presented below each
structure, with the value for the most stable configuration
highlighted in red.
\label{figA1}}
\end{figure}

\begin{figure}
\includegraphics[width=0.8\columnwidth]{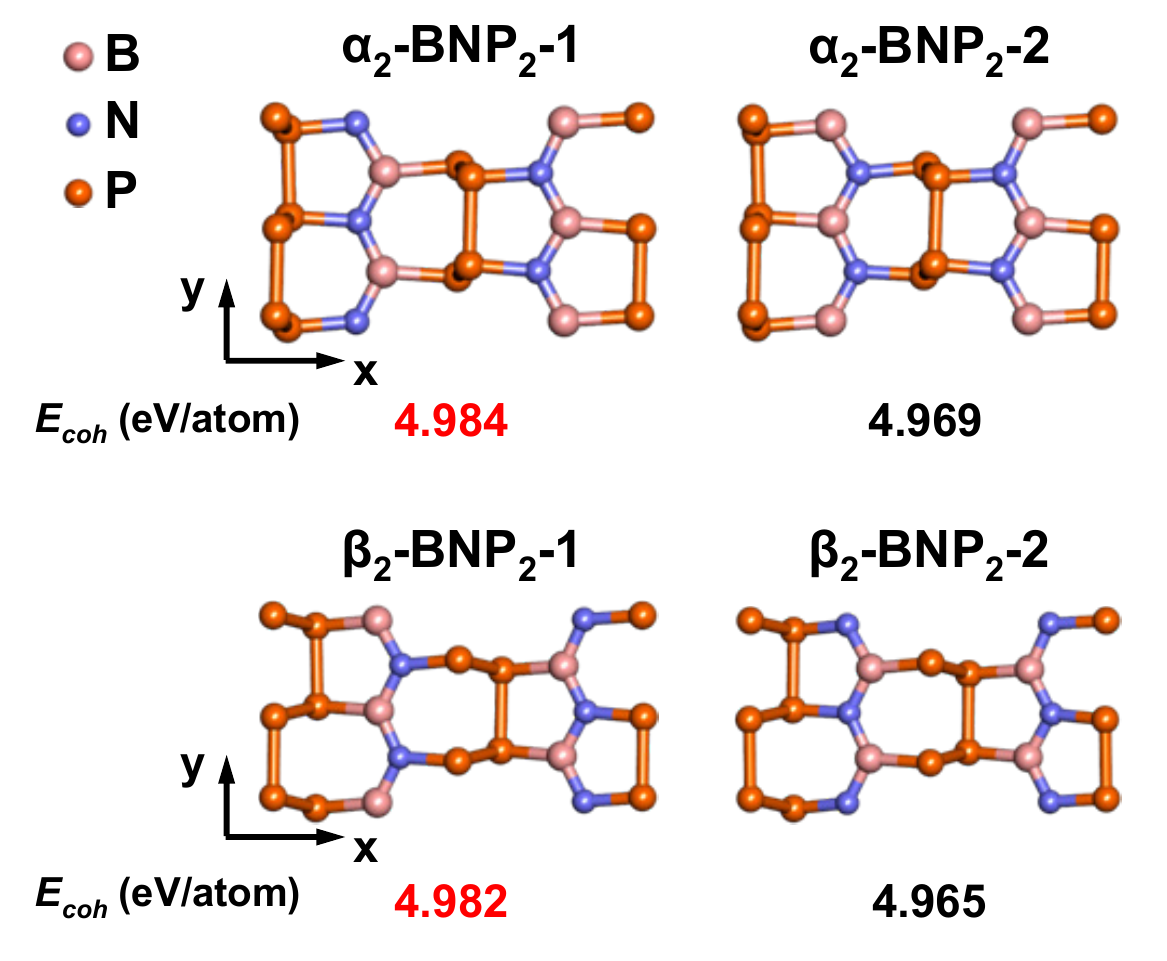}
\caption{Ball-and-stick model of metastable structures related to
the $\alpha_1$ and $\beta_1$ allotropes of C$_2$SiS, shown in top
view. Cohesive energies $E_{coh}$ are presented below each
structure, with the value for the most stable configuration
highlighted in red.
\label{figA2}}
\end{figure}

\begin{figure}
\includegraphics[width=0.9\columnwidth]{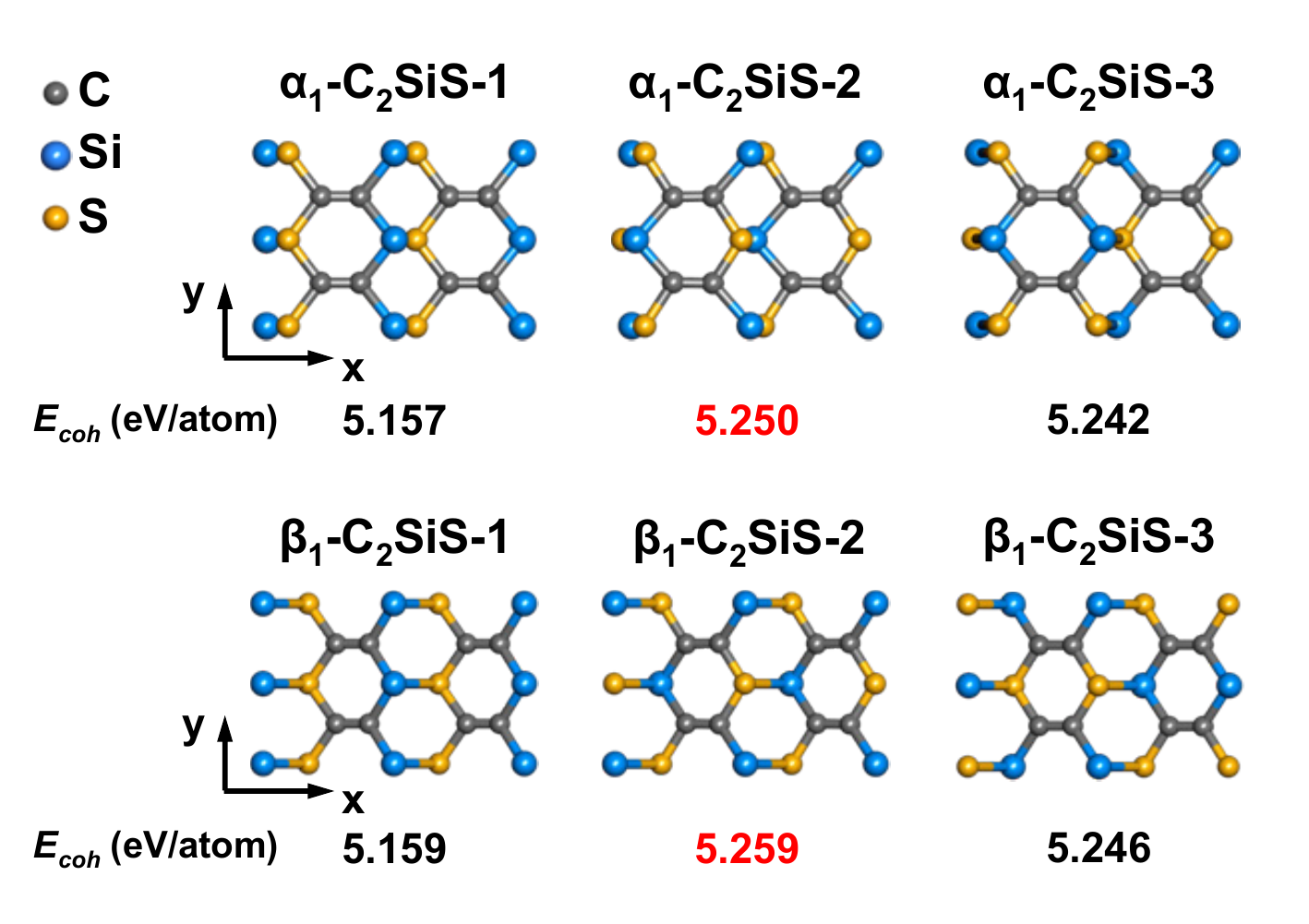}
\caption{Ball-and-stick model of metastable structures related to
the $\alpha_2$ and $\beta_2$ allotropes of BNP$_2$, shown in top
view. Cohesive energies $E_{coh}$ are presented below each
structure, with the value for the most stable configuration
highlighted in red.
\label{figA3} }
\end{figure}

\begin{figure*}[t]
\includegraphics[width=1.8\columnwidth]{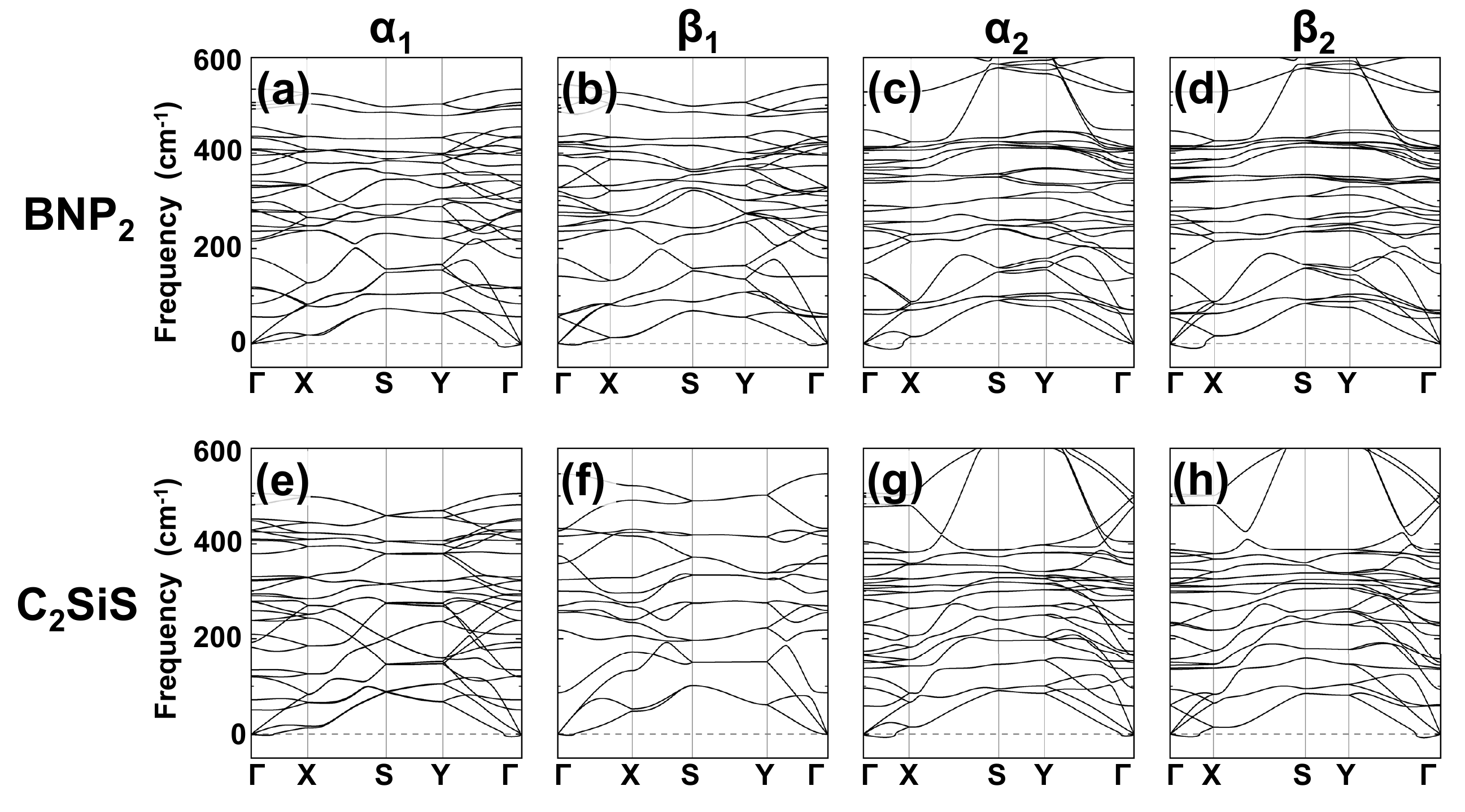}
\caption{Phonon spectra for monolayers of %
(a) $\alpha_1$-BNP$_2$,  %
(b) $\beta_1$-BNP$_2$,   %
(c) $\alpha_2$-BNP$_2$,  %
(d) $\beta_2$-BNP$_2$,   %
(e) $\alpha_1$-C$_2$SiS, %
(f) $\beta_1$-C$_2$SiS,  %
(g) $\alpha_2$-C$_2$SiS, and %
(h) $\beta_2$-C$_2$SiS.
\label{figA4}}
\end{figure*}

\begin{figure*}
\includegraphics[width=1.7\columnwidth]{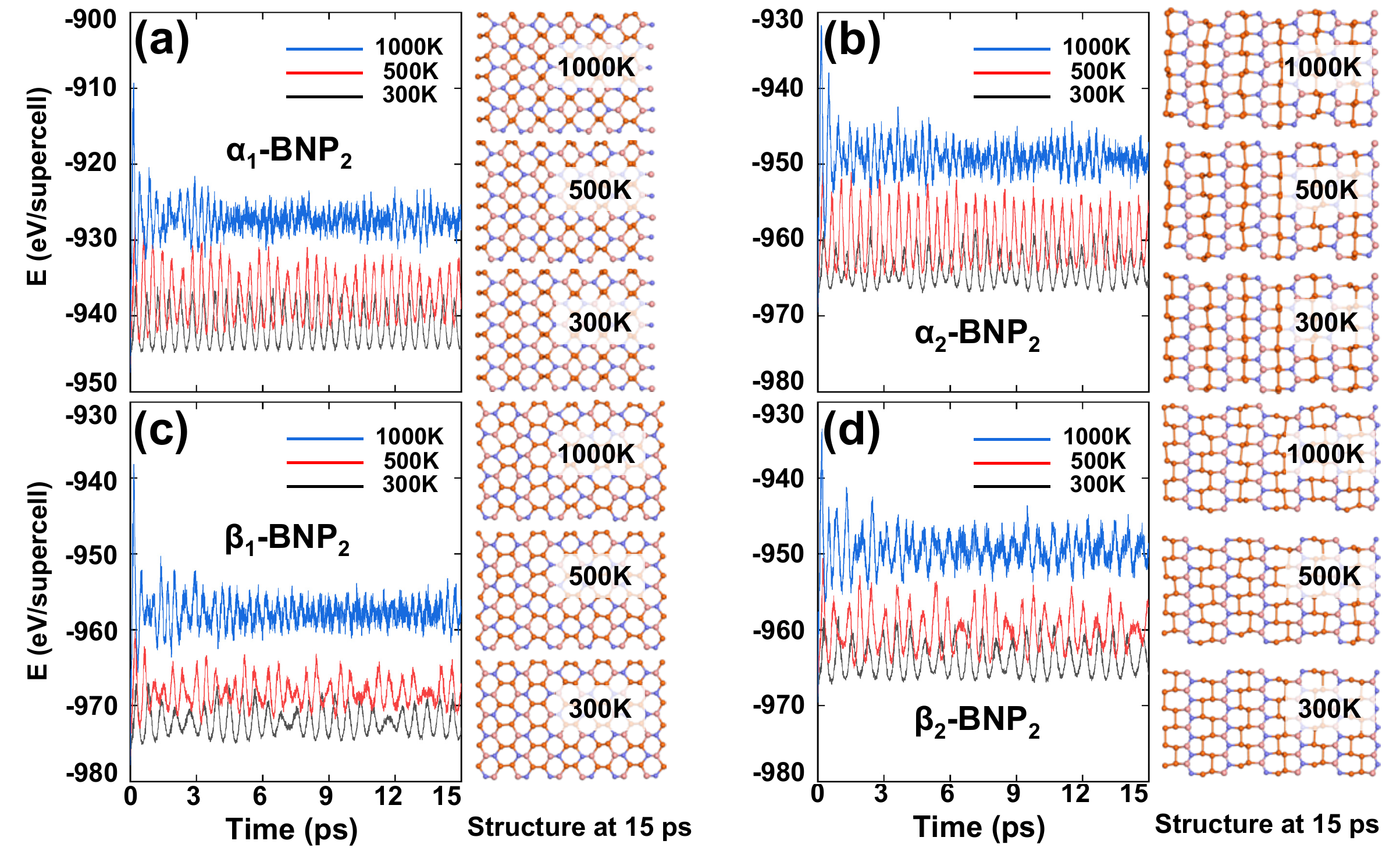}
\caption{Fluctuations of the total potential energy (left panels)
and structural snap shots after $15$~ps (right panels) for %
(a) $\alpha_1$-BNP$_2$, %
(b) $\alpha_2$-BNP$_2$, %
(c) $\beta_1$-BNP$_2$, and %
(d) $\beta_2$-BNP$_2$ %
monolayers during canonical MD simulations at $300$~K, $500$~K and
$1000$~K.
\label{figA5}}
\end{figure*}

\begin{figure*}[t]
\includegraphics[width=1.7\columnwidth]{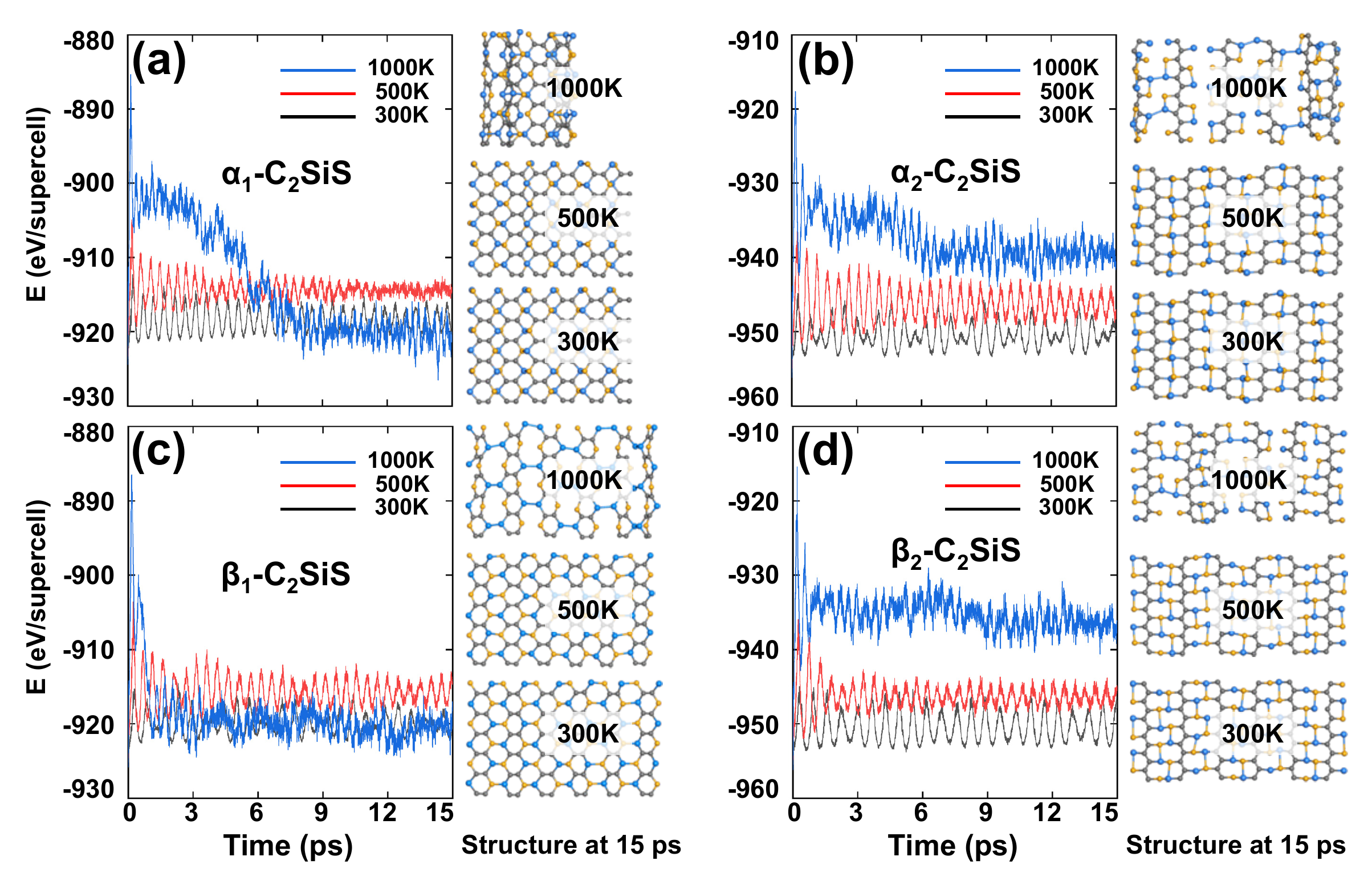}
\caption{Fluctuations of the total potential energy (left panels)
and structural snap shots after $15$~ps (right panels) for %
(a) $\alpha_1$-C$_2$SiS, %
(b) $\alpha_2$-C$_2$SiS, %
(c) $\beta_1$-C$_2$SiS, and %
(d) $\beta_2$-C$_2$SiS %
monolayers during canonical MD simulations at $300$~K, $500$~K and
$1000$~K.
\label{figA6}}
\end{figure*}

\begin{figure*}
\includegraphics[width=1.0\columnwidth]{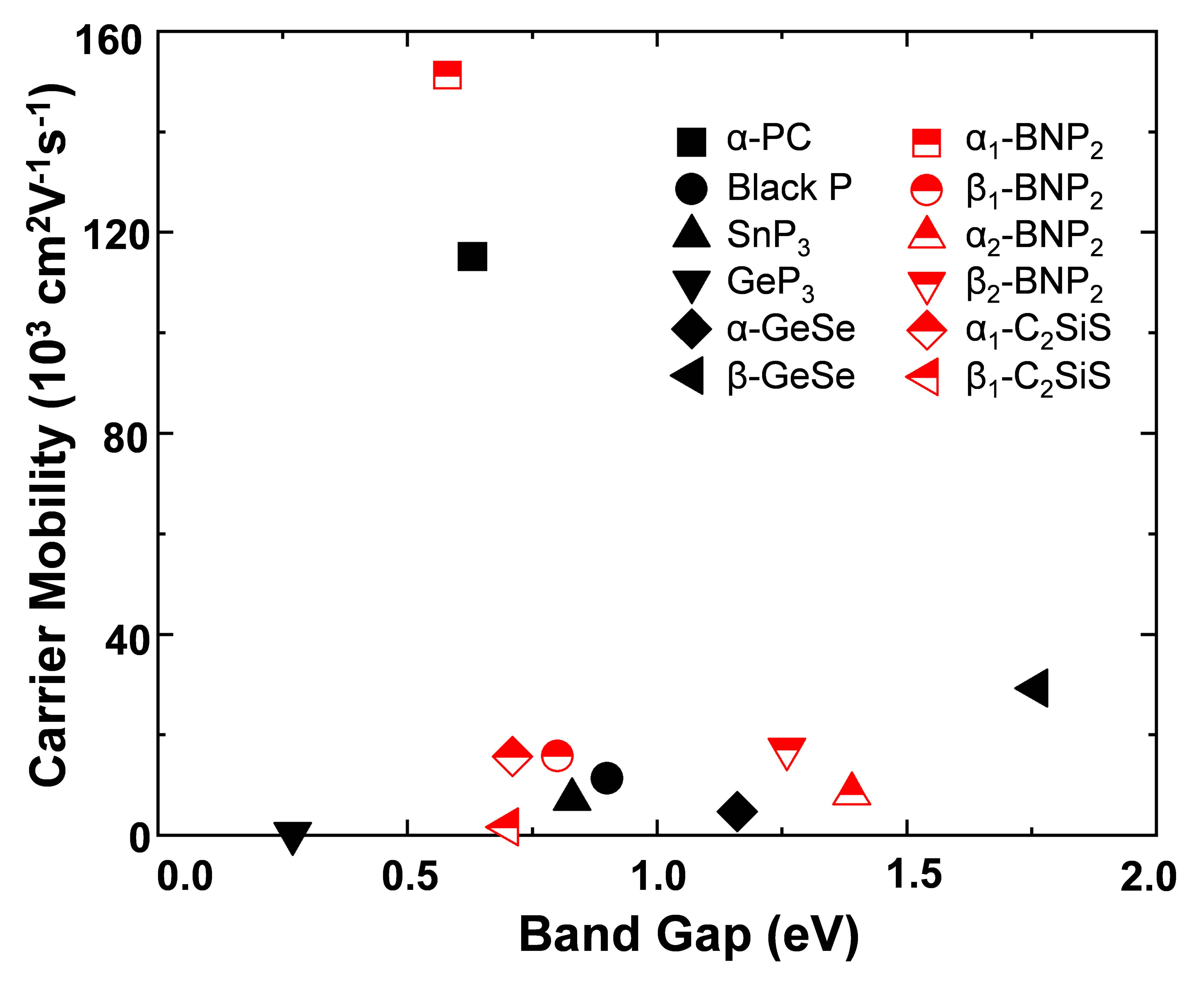}
\caption{%
\modR{ %
Carrier mobility versus band gap for different 2D materials.
Results for semiconducting allotropes predicted in this study,
shown by the red symbols, are compared to previously reported
values for selected 2D materials, shown by the black symbols. All
results have been obtained at the PBE level of DFT.
}%
\label{figA7}}
\end{figure*}

As seen in Fig.~\ref{fig1} of the main text, $\alpha_1$ and
$\beta_1$ allotropes of BNP$_2$ contain four polar B-N dimers
aligned with the $x$-axis in each unit cell. Each of these dimer
bonds may be rotated by 180$^\circ$. This gives rise to four
metastable structures for $\alpha_1$-BNP$_2$ and another four
structures for $\beta_1$-BNP$_2$, shown in Fig.~\ref{figA1}. The
most stable structures among these, with the cohesive energy value
highlighted in red, have been considered in the main text.

Quite different is the arrangement of B and N atoms in zigzag
chains in the $\alpha_2$ and $\beta_2$ allotropes of BNP$_2$ shown
in Fig.~\ref{fig1} of the main text. Also in this case, each bond
may be rotated by 180$^\circ$. This gives rise to two metastable
structures for $\alpha_2$-BNP$_2$ and another two structures for
$\beta_2$-BNP$_2$, shown in Fig.~\ref{figA2}. Only the most stable
among these allotropes have been considered in the main text.

Next we turn to the $\alpha_1$ and $\beta_1$ allotropes of
C$_2$SiS, displayed in Fig.~\ref{fig1}(e) and (f) of the main
text. Unlike in BNP$_2$, we can find only three inequivalent
metastable structures for these systems, displayed in
Fig.~\ref{figA3}. For the $\alpha_2$ and $\beta_2$ allotropes of
C$_2$SiS, optimization leads to only one stable structure. As for
the other systems discussed above, we only consider only the most
stable allotropes in the main text.

\subsection*{Phonon spectra of 2D BNP$_2$ and C$_2$SiS allotropes}

A real confirmation of structural stability, more important than a
high cohesive energy, comes from the phonon spectra. Structures
can be considered stable if no imaginary frequencies can be found
in the phonon spectra. The calculated phonon spectra of all
BNP$_2$ and C$_2$SiS allotropes discussed in this study are
displayed in Fig.~\ref{figA4}.

We should note at this point that phonon calculations for 2D structures
with the uniquely soft flexural ZA mode are very demanding on
precision%
\modR{%
~\cite{2dphonon}. We found our phonon spectra to be essentially
free of imaginary frequencies associated with decay modes. The
highest imaginary frequency value
${\omega}{\lesssim}12i$~cm$^{-1}$ in our spectra belongs to the ZA
mode near the $\Gamma$ point in the Brillouin zone and is an
artifact related to numerical precision~\cite{2dphonon}.
}%

\subsection*{Thermodynamic stability of 2D BNP$_2$ and
             C$_2$S\lowercase{i}S allotropes}

Whereas phonon spectra tell about structural stability in the harmonic
regime, they cannot determine if a structure will or will not fall
apart at a given finite temperature. To study the thermodynamic
stability of the ternary structures in this study, we performed a set
of canonical {\em ab initio} molecular dynamics (MD) simulations at
$300$~K, $500$~K and $1000$~K and present our results in
Fig.~\ref{figA5} for 2D BNP$_2$ and Fig.~\ref{figA6} for C$_2$SiS
structures. For $15$~ps long MD runs, we plotted both the fluctuations
of the total potential energy and snapshots of the structures after
$15$~ps.

Our results in Fig.~\ref{figA5} indicate that all four allotropes
of BNP$_2$ maintained their geometries up $1000$~K, indicating a
high thermodynamic stability. The corresponding results for
C$_2$SiS in Fig.~\ref{figA6} indicate all allotropes to be stable
at $300$~K and $500$~K. Further increase of temperature to
$1000$~K causes a dramatic degradation of the $\alpha_1$ and
$\beta_1$ allotropes of C$_2$SiS after $3$~ps. The $\alpha_2$ and
$\beta_2$ allotropes appear more stable at $1000$~K, but their
significant distortion indicates onset of degradation.
\modR{ %
The thermodynamic stability of all the BNP$_2$ and C$_2$SiS
allotropes is either comparable to or superior to other similar 2D
structures including GeP$_3$~\cite{Jing17GeP3}, which was shown to
be stable at $T{\lesssim}500$~K, as well as PC~\cite{Wang16PC} and
binary V-V compounds~\cite{Yu16jmcc}, which were shown to be
stable only up to ${\approx}300$~K. Our MD results show good
consistency with the conclusions obtained from the formation
energy calculations discussed in the main text.
}%

\subsection*{%
\modR{ %
Suitability of BNP$_2$ and C$_2$S\lowercase{i}S allotropes for
electronic applications %
}%
}

\modR{ %
Combination of high carrier mobility and wide band gap is very
desirable, as it results in a high ON/OFF ratio and a high
on-state current in semiconductor devices. The 2D materials we
introduce appear particularly suitable for electronic applications
from that viewpoint. This is evidenced in Fig.~\ref{figA7}, where
we compare mobilities and band gaps of BNP$_2$ and C$_2$SiS to
those of 2D-PC, GeP$_3$~\cite{Jing17GeP3}, SnP$_3$~\cite{SnP3},
black phosphorene, and 2D GeSe~\cite{GeSe17}. There is a general
trade-off relation between carrier mobility and band gap. We find
most of the BNP$_2$ and C$_2$SiS allotropes to be near the upper
limit of the trade-off.
}%



%

\end{document}